\documentclass[submitting]{nst}

\usepackage{subfigure,dcolumn}
\usepackage[T2A,T1]{fontenc}
\usepackage[russian,english]{babel}
\usepackage{hyperref}
\usepackage{url}
\newcommand{\tabincell}[2]{\begin{tabular}{@{}#1@{}}#2\end{tabular}}
\allowdisplaybreaks

\usepackage{xcolor, soul} 
\soulregister{\cite}{7}
\soulregister{\ref}{7}
\soulregister{\bibitem}{7}
\soulregister{\href{}}{7}

\sethlcolor{yellow}

\sethlcolor{yellow}

\usepackage{makecell}

\usepackage{listings}
\begin{document}

\title{Extraction of fissile isotope antineutrino spectra using feedforward neural network}
\thanks{We thank Yuda Zeng for the fruitful discussions. This work was supported by the China Postdoctoral Science Foundation under Grant Number 2024M753715, Fundamental Research Funds for the Central Universities, Sun Yat-sen University (Grant Nos. 24qnpy125 and 22lglj11), and Guangdong Basic and Applied Basic Research Foundation (Grant No. 2023B1515120030).}

\author{Jian Chen}
\affiliation{School of Physics and Astronomy, Sun Yat-sen University, Zhuhai, 519082, China}
\author{Jun Wang}
\affiliation{Sino-French Institute of Nuclear Engineering and Technology, Sun Yat-sen University, Zhuhai, 519082, China}
\email[Corresponding authors, ]{Jun Wang, wangj933@mail.sysu.edu.cn\\Wei Wang, wangw223@mail.sysu.edu.cn\\Yuehuan Wei, weiyh29@mail.sysu.edu.cn}
\author{Wei Wang}
\affiliation{Sino-French Institute of Nuclear Engineering and Technology, Sun Yat-sen University, Zhuhai, 519082, China}
\affiliation{School of Physics, Sun Yat-sen University, Guangzhou, 510275, China}
\author{Yuehuan Wei}
\affiliation{Sino-French Institute of Nuclear Engineering and Technology, Sun Yat-sen University, Zhuhai, 519082, China}

\begin{abstract}
 The precise measurement of the antineutrino spectra produced by isotope fission in reactors is of great significance for studying neutrino oscillations, refining nuclear databases, and addressing the reactor antineutrino anomaly. In this paper, we report a method that utilizes a feedforward neural network (FNN) model to decompose the prompt energy spectrum observed in a short-baseline reactor neutrino experiment and extract the antineutrino spectra produced by the fission of major isotopes such as $^{235}$U, $^{238}$U, $^{239}$Pu, and $^{241}$Pu in the nuclear reactor. We present two training strategies for the model and compare them with the traditional $\chi^2$ minimization method by applying them to the same set of pseudo-data corresponding to a total exposure of $(2.9\times 5\times 1800)~\rm{GW_{th}\cdot tonnes\cdot days}$. The results show that the FNN model not only converges faster and better during the fitting process but also achieves relative errors of less than 1\% in the $2-8$ MeV range in the extracted spectra, outperforming the $\chi^2$ minimization method. The feasibility and superiority of this method were validated in the study.
\end{abstract}

\keywords{Reactor neutrinos, Isotope antineutrino spectra, Feedforward neural network }

\maketitle

\section{Introduction}
\label{sec:intro}

Since the direct discovery of neutrinos by Cowan and Reines  at the Savannah River reactor power plant in 1956~\cite{reines1960detection}, reactor neutrino experiments have played a pivotal role in the advancement of neutrino physics. Reactor neutrinos are also known as reactor antineutrinos because they are composed exclusively of electron antineutrinos ($\bar{\nu}_e$). In commercial pressurized water reactors (PWRs), more than 99.7\% of the reactor neutrinos are emitted from the beta decay branches of neutron-rich fission products generated by four isotopes: $^{235}$U, $^{238}$U, $^{239}$Pu, and $^{241}$Pu. In research reactors utilizing 93\% $^{235}$ U-enriched fuel, 99.3\% of the reactor neutrinos result from the fission of $^{235}$U. Each isotope releases approximately six $\bar{\nu}_e$ per fission along with a corresponding antineutrino flux and spectrum. Precise fissile isotope antineutrino spectra are required for reactor monitoring and safeguarding applications ~\cite{bernstein2002nuclear,fallot2019antineutrino,Bowden:2022rjt} and serve as valuable inputs to reactor neutrino experiments utilizing the inverse beta decay (IBD) reaction~\cite{JUNO:2015zny,NEOS:2016wee,PROSPECT:2020sxr} or coherent elastic neutrino-nucleus scattering (CE$\nu$NS)~\cite{CONUS:2020skt,RELICS:2024opj}.

Fissile isotope antineutrino spectra and fluxes have been evaluated several times in the past decades. The methodologies employed can be classified into three major categories comprising summation ~\cite{Fallot:2012jv}, conversion ~\cite{Schreckenbach:1981wlm, Mention:2011rk}, and extraction methods ~\cite{DayaBay:2019yxq,DayaBay:2021dqj,DayaBay:2021owf}. The summation method, i.e., the \textit{ab initio} approach, utilizes information on fission products and decays from nuclear databases to calculate and sum the contributions of all possible beta decay chains to $\bar{\nu}_e$~\cite{Mueller:2011nm}. 
However, the presence of incomplete or inaccurate information in nuclear databases introduces complexities and challenges in constructing reliable spectral models, ultimately leading to potentially large and unknown uncertainties in model predictions. The conversion method relies on the measured beta spectra of uranium and plutonium. The beta spectra for thermal-neutron-induced fissions of $^{235}$U, $^{239}$Pu, and $^{241}$Pu have been measured at the Institut Laue-Langevin High Flux Reactor in the 1980s ~\cite{Schreckenbach:1981wlm, VonFeilitzsch:1982jw, Schreckenbach:1985ep, Hahn:1989zr}, while those for the fast-neutron-induced fission of $^{238}$U were measured at the Heinz Maier-Leibnitz (FRM II) research neutron source in 2013~\cite{Haag:2013raa}. The measured beta spectra for each isotope are fitted by a set of virtual beta decay branches based on the allowed beta decay transitions, which are then converted into antineutrino branches and summed to the corresponding isotope antineutrino spectra~\cite{Mueller:2011nm, PhysRevC.84.024617}. Although the reduced dependence on nuclear databases in this method provides spectral shapes with typical relative uncertainties of a few percent, the fine structural information in the spectral shapes is not as rich as that obtained using the summation method. 
To address these shortcomings, several antineutrino spectrum models have been developed based on the conversion method or a combination of both methods. One example is the Huber-Mueller model~\cite{Mueller:2011nm, PhysRevC.84.024617}, which provides predictions that roughly agree with earlier experimental data and is widely accepted in reactor neutrino experiments. 
However, measurements from short-baseline reactor neutrino experiments such as Double Chooz~\cite{DoubleChooz:2019qbj}, RENO~\cite{RENO:2018dro}, Daya Bay~\cite{DayaBay:2019yxq}, and NEOS~\cite{NEOS:2016wee} confirmed a $\sim6\%$ deficit in the measured reactor antineutrino flux and an excess in the $4-6$ MeV prompt energy range compared to the predictions of the Huber-Mueller model.
These discrepancies, which are respectively known as the ``reactor antineutrino anomaly (RAA)"~\cite{Mention:2011rk} and ``5 MeV excess" or ``5 MeV bump"~\cite{Seo:2014xei,Huber:2016xis}, cannot be ignored in the era of precise measurements. The extraction method, in which the fission isotope antineutrino spectrum is inferred from the reconstructed prompt energy spectrum measured by the detector and independent of nuclear databases, has become a common approach for testing various RAA formation hypotheses, including explanations of sterile neutrinos.
Using this method, the Daya Bay experiment \cite{DayaBay:2019yxq,DayaBay:2021dqj} extracted the 235U and 239Pu antineutrino spectra from PWRs, while the PROSPECT \cite{DayaBay:2021owf,PROSPECT:2022wlf} and STEREO \cite{Stereo:2021wfd,STEREO:2022nzk} experiments extracted the 235U antineutrino spectrum from highly enriched uranium research reactors. 
Moreover, it was revealed that the flux deficit was primarily carried by $^{235}$U, and the 5 MeV bump had shared contributions from uranium and plutonium. However, the extraction of the $^{238}$U and $^{241}$Pu antineutrino spectra was not satisfactory owing to statistical limitations ~\cite{DayaBay:2021dqj}. 

The current general practice in experiments for extracting fissile isotope antineutrino spectra involves first unfolding the reconstructed prompt energy spectrum to obtain an antineutrino energy spectrum weighted by the IBD cross section, and then further fitting the unfolded spectrum with the $\chi^2$ minimization method to extract individual or combined isotope antineutrino spectra~\cite{DayaBay:2021dqj,DayaBay:2021owf,Stereo:2021wfd, PROSPECT:2022wlf}. 
Unfolding is a common technique used in high-energy physics (HEP) to disentangle detector effects, correct migration effects, suppress fluctuations, and reconstruct approximate distributions of quantities. Common methods for unfolding include singular value decomposition (SVD)~\cite{Hocker:1995kb}, Wiener SVD ~\cite{Tang:2017rob}, and Bayesian iterations ~\cite{DAgostini:1994fjx}. In the Daya Bay experiment, these methods were used to yield consistent extraction results. Although the Wiener-SVD method produces the smallest unfolded spectrum mean square error (MSE) within the energy range of $3-6$ MeV,  it does not perform as well as the other methods outside this energy range because of the large statistical fluctuations in the intrinsic neutrino energy spectrum~\cite{DayaBay:2021dqj}. To obtain more precise solutions, the number of bins for the unfolded spectrum in experiments is typically limited to that of the intrinsic spectrum~\cite{DayaBay:2016ssb}. Although this simplifies the subsequent fitting process for extracting the specific fission isotope antineutrino spectrum, it also suppresses the fine structure of the spectrum shape.

In our previous study ~\cite{Zeng:2023att}, we proposed a machine learning method in which a convolutional neural network (CNN) model is employed to extract fission isotope antineutrino spectra from the unfolded prompt energy spectrum in a virtual short-baseline reactor neutrino experiment. The analysis results demonstrate that the proposed CNN model can achieve subpercentage uncertainties in the extracted $^{235}$U and $^{239}$Pu antineutrino spectra whereas the $^{238}$U and $^{241}$Pu antineutrino spectra need to be constrained via prior knowledge during the fitting process. In this study, we extend the method and establish a feedforward neural network (FNN) model to resolve this extraction problem. This new method is designed to directly extract the antineutrino spectra of the four fission isotopes from the reconstructed prompt energy spectrum without highlighting the unfolding process or any constraints on the spectra while better preserving the fine structure of the extracted spectra.

The remainder of this paper is organized as follows: In Section ~\ref{sec:general_description}, we present the antineutrino spectra of the IBD reactions and the generation of the simulation dataset for this study. In Section~\ref{sec:ffnn}, we introduce the conceptual and technical details of the proposed FNN model and its training strategies. In Section ~\ref{sec:analysis&results}, we compare the performance of this new method in extracting fission isotope antineutrino spectra with that of the benchmark traditional method, that is, the $\chi^2$ minimization method, and discuss the obtained results. Finally, a summary and future outlook are presented in Section ~\ref{sec:summary&outlook}.

\section{Dataset generation for FNN model}

In this study, we constructed a virtual reactor neutrino experiment in a layout comprising a PWR and a detector. To verify the feasibility of the virtual experiment, we referred to the Daya Bay~\cite{DayaBay:2021dqj} and Taishan Antineutrino Observatory (TAO, also known as JUNO-TAO)~\cite{JUNO:2020ijm,luo2023design} experiments, and made the following assumptions about the experimental parameters: The reactor is operated for 1800 days at a full thermal power of 2.9~$\mathrm{GW_{th}}$ with an initial uranium fuel mass of 72 tons. The detector is loaded with 5 tonnes of liquid scintillator (LS) with 12\% hydrogen by mass, has an energy resolution of 8\% at 1 MeV and detection efficiency of 50\%, and is situated at a baseline distance of 30 m. We adopted the Huber-Mueller model as the foundational theory for the phenomenological prediction of the IBD yield to generate the simulated sample dataset for this study. The model selection did not significantly affect the analysis. We disregarded the contributions of the spent nuclear fuel and the non-equilibrium effect on the IBD yield~\cite{Mueller:2011nm,DayaBay:2016ssb}.

\label{sec:general_description}
\subsection{IBD yield prediction}
The Huber-Mueller model is a theoretical framework for predicting the antineutrino spectra produced by the fission reactions of four isotopes in reactors. Each of these isotopic antineutrino spectrum can be parameterized using the exponent of a fifth-order polynomial as follows:
\begin{equation} 
    s_{l}(E_\nu)=\exp\left(\sum_{p=1}^6\alpha_{lp}E_\nu^{p-1}\right),
     \label{eq:iso_antineu_spectrum}
\end{equation}
where $l=\{^{235}\rm{U},~^{238}\rm{U},~^{239}\rm{Pu},~^{241}\rm{Pu}\}$, $E_\nu$ is the $\bar{\nu}_{e}$ energy, and the $\alpha_{lp}$s are polynomial coefficients for the isotope $l$. The $\alpha_{lp}$ coefficients for $^{235}$U, $^{239}$Pu, and $^{241}$Pu were derived using the conversion method by Huber~\cite{PhysRevC.84.024617}, whereas the $\alpha_{lp}$ coefficients for $^{238}$U were obtained using the summation method of Mueller et al. ~\cite{Mueller:2011nm}. To incorporate the RAA in this study, we modified the isotopic antineutrino spectrum in Eq. ~(\ref{eq:iso_antineu_spectrum}) as follows: 
\begin{equation} 
    S_{l}(E_\nu)=s_{l}(E_\nu)r_{RAA}(E_\nu), 
     \label{eq:raa_iso_antineu_spectrum}
\end{equation}
where $r_{RAA}(E_\nu)$ is the ratio of the RAA between the spectra measured in the Daya Bay experiment~\cite{DayaBay:2016ssb} and the Huber-Mueller model prediction. To evaluate $r_{RAA}(E_\nu)$, we performed cubic spline interpolation within the provided energy range of $1.8-8$ MeV and set it uniformly to 1 for energy values above 8 MeV.

The antineutrino yield per fission can be expressed as
\begin{equation} 
    \phi(E_{\nu},t)=\sum_lf_l(t)S_l(E_\nu),
    \label{antineutrinosyield}
\end{equation}
where the fission fraction $f_l(t)$ represents the relative contribution of the isotope $l$ to the fission reaction at time $t$. 
The event rate of antineutrinos emitted from the reactor core can be calculated as
\begin{equation} 
    \frac{dN}{dE_\nu}=\frac{W(t)}{\sum\limits_lf_l(t)\epsilon_l}\phi(E_{\nu},t),
    \label{eventrateofantineutrinos}
\end{equation}
where $W(t)$ is the thermal power of the reactor at time $t$, $\epsilon_l$ is the mean energy released per fission of the isotope $l$, and the values for $\epsilon_l$ were obtained from Ref.~\cite{Ma:2012bm}.

In the standard three-flavor neutrino oscillation framework, the survival probability  $P_{ee}$ of $\bar{\nu}_{e}$ after propagating a distance $L$ is given by~\cite{Workman:2022ynf}
\begin{equation}
\begin{aligned}
P_{e e}\left(L, E_{\nu}\right)&= P\left(\bar{\nu}_{e} \rightarrow \bar{\nu}_{e} ; L, E_{\nu}\right) \\
&= 1 - \cos ^{4} \theta_{13} \sin ^{2}\left(2 \theta_{12}\right) \sin ^{2}\left(\Delta_{21}\right) \\
& \phantom{= 1\ } - \cos ^{2} \theta_{12} \sin ^{2}\left(2 \theta_{13}\right) \sin ^{2}\left(\Delta_{31}\right) \\
& \phantom{= 1\ } - \sin ^{2} \theta_{12} \sin ^{2}\left(2 \theta_{13}\right) \sin ^{2}\left(\Delta_{32}\right),
\end{aligned}
\label{survivalprobability}
\end{equation}
where the $\theta_{ij}$s represent the neutrino mixing angles. The oscillation phases $\Delta_{ij}$ are given by 
\begin{equation}
\begin{aligned}
\Delta_{ij}=\frac{\Delta m_{ij}^2L}{4E_\nu}\simeq\frac{1.267\Delta m_{ij}^2[\mathrm{eV^2}]L[\mathrm{m}]}{E_\nu[\mathrm{MeV}]}, 
\end{aligned}
\label{oscillationphases}
\end{equation}
where  $\Delta m_{ij}^2$ denotes the mass-squared difference between the two mass eigenstates $m_i$ and $m_j$, i.e., $\Delta m_{ij}^2\equiv m_i^2-m_j^2$.

For short-baseline reactor neutrino experiments, considering that the term involving $\Delta_{21}$ is negligible and $\Delta m_{31}^2\approx\Delta m_{32}^2$, Eq. (\ref{survivalprobability}) can be simplified to

\begin{equation}
\begin{aligned}
P_{ee}\left(L, E_{\nu}\right)\approx1-\sin^2(2\theta_{13})\sin^2(\frac{\Delta m_{31}^2L}{4E_\nu}).
\end{aligned}
\label{neutrinosurvival_SL}
\end{equation}
Unless otherwise specified,  $\sin^{2}\theta_{13}= (2.20±0.07)×10^{-2}$, $\Delta\text{m}_{32}^2=(2.437\pm0.033)\times10^{-3}\text{ eV}^2$, and $\Delta\text{m}_{21}^2=(7.53\pm0.18)\times10^{-5}\text{ eV}^2$ in this study based on the values from the Particle Data Group (PDG) 2022~\cite{Workman:2022ynf}.

As the $\bar{\nu}_e$ emitted by the reactor propagate to the LS detector, some of them engage in IBD reactions with the free target protons in the LS, which are denoted as $\bar{\nu}_e+p\rightarrow e^++n$. In this process, the positron $e^+$ rapidly deposits its energy and annihilates the surrounding electron $e^-$ to form two 0.511 MeV gammas, generating a prompt signal. The neutron $n$ scatters within the detector until it is thermalized and subsequently captured by hydrogen (99\%) or carbon (1\%) within $\sim200~\mu s$, thereby releasing a 2.22 or 4.95 MeV gamma, respectively, and yielding a delayed signal~\cite{JUNO:2020xtj}. 
An IBD event is identified by the prompt-delayed signal pair during such a brief interval. 
The measured IBD event number $M_k$ in the $k$-th reconstructed prompt energy $E_{rec}$ bin observed at a detector within the data acquisition time $T_{DAQ}$ is therefore given by~\cite{JUNO:2015zny}

\begin{equation}
\begin{aligned}
\mathrm{M}_{k}&=\frac{N_{p}\varepsilon}{4\pi L^{2}}\int\limits_{E_{rec}^{k}}^{E_{rec}^{k+1}}dE_{rec}\int\limits_{T_{DAQ}}dt\int\limits_{E_{thr}}dE_{\nu}\\&\times\frac{dN}{dE_{\nu}}P_{ee}(L,E_{\nu})\sigma_{IBD}(E_{\nu})G(E_\nu,E_{rec})\\
\end{aligned}
\label{ObsAnt_Eq}
\end{equation}
where $N_p$ is the number of free target protons in the LS, $\varepsilon$ is the detection efficiency of the detector, the IBD threshold energy $E_{thr}\sim m_n-m_p+m_e\sim1.8$ MeV, $\sigma_{IBD}(E_\nu)$ is the cross-section of the IBD taken from Ref. ~\cite{Strumia:2003zx}, and $G(E_{rec}, E_\nu)$ is a normalized Gaussian smearing function, which includes the energy resolution effect.

\begin{figure}[ht]
    \centering
    \includegraphics[width=0.5\textwidth]{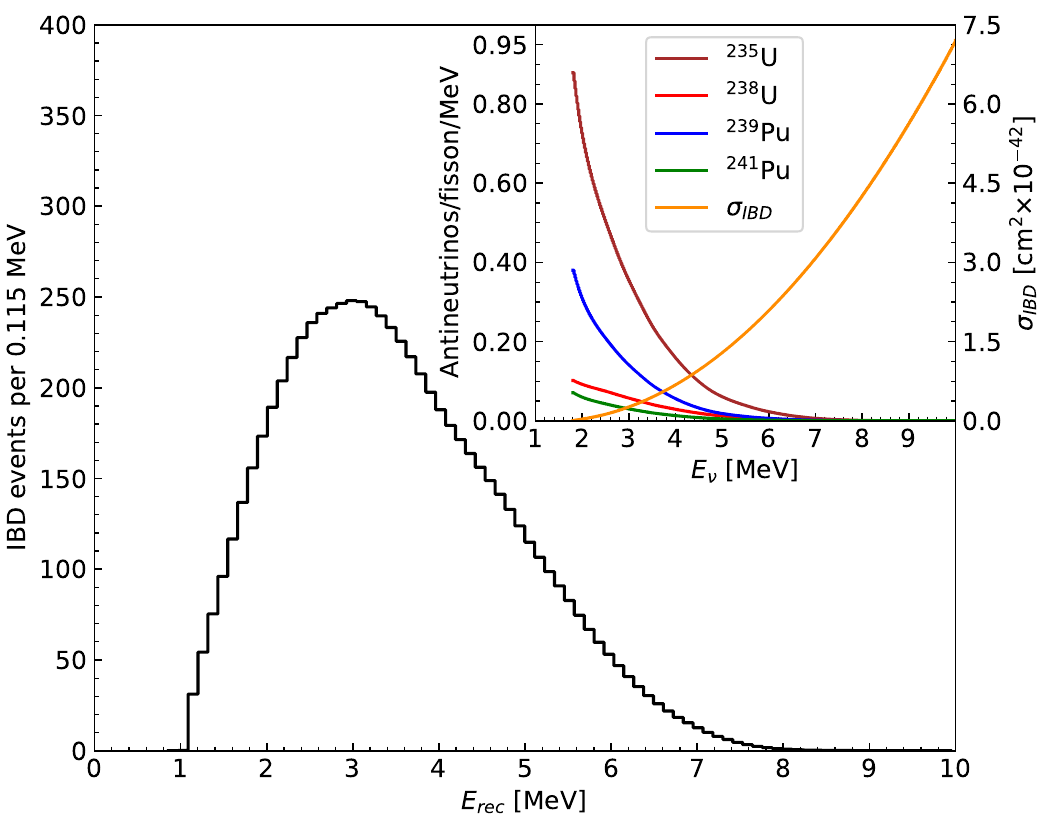}
    \caption{The virtual detector observes the reconstructed prompt energy spectrum distorted by the RAA with an exposure of $(2.9\times5\times1)~\rm{GW_{th}\cdot tonnes\cdot day}$. The insert shows the cross-section of the IBD reaction~\cite{Strumia:2003zx} and the four isotope antineutrino spectra obtained by modifying the Huber-Muller model according to Eq.~(\ref{eq:raa_iso_antineu_spectrum}).}
   \label{fig:phi_IBDCrossFission} 
\end{figure}

To simplify the calculation, we assumed that the detector has no energy leakage or LS nonlinearity ~\cite{DayaBay:2016ssb}. Thus, the prompt energy $E_{pro}\sim E_\nu-0.78$ MeV, and $E_{rec}$ is expected to obey the distribution $G(E_{rec}, E_\nu)$ defined as follows~\cite{JUNO:2015zny}:
\begin{equation}
\begin{aligned}
G(E_\nu,E_{rec})\simeq\frac{1}{\sqrt{2\pi}\delta_{E_{pro}}}\exp\left\{-\frac{\left(E_{pro}-E_{rec}\right)^{2}}{2(\delta_{E_{pro}})^{2}}\right\}, 
\end{aligned}
\label{GaussianF_Eq}
\end{equation}
The energy resolution $\delta_{E_\mathrm{pro}}$ is parameterized
\begin{equation}
\begin{aligned}
\frac{\delta_{E_\mathrm{pro}}}{E_{pro}}=\sqrt{\left(\frac {p_0}{\sqrt{E_{pro}}}\right)^2+p_1^2+\left(\frac {p_2}{E_{pro}}\right)^2}, 
\end{aligned}
\label{Denergyresolution_Eq}
\end{equation}
where $p_0$ quantifies the statistical fluctuations in the photons detected by the detector, $p_1$ is predominantly influenced by residual effects resulting from the spatial nonuniformity and temporal instability correction of the detector, and $p_2$ quantifies the effects associated with the photomultiplier tube (PMT), notably the PMT dark noise \cite{JUNO:2015zny,JUNO:2020xtj}.

For simplicity, we set $p_0 = 0.08$, $p_1 = 0$, and $p_2 = 0$ in this study for an energy resolution of 8\% at 1 MeV in the detector. Therefore, under full reactor power and classical fission fractions conditions ~\cite{DayaBay:2016ssb}, the detector observes the energy spectrum of the IBD events (i.e., the reconstructed prompt energy spectrum) distorted by the RAA in one day, as shown in figure~\ref{fig:phi_IBDCrossFission}, and approximately 7473 IBD events are recorded.

\subsection{Simulated samples and targets in dataset}
\label{sebsec:samples}

Considering the significant computational resources and time required for the integral terms in Eq. (\ref{ObsAnt_Eq}), Eq. (\ref{ObsAnt_Eq}) is typically converted to a discrete summation or matrix multiplication equivalent form in practical computations. In this study, the integral form of the reconstructed prompt energy spectrum is rewritten as an element of the row matrix $M_{1\times N_{E_{rec}}}$, which is given in Eq. (\ref{matrix_ibd_spec}). Each element of the matrix represents the measured IBD event number in the corresponding energy bin.
\begin{widetext}
\begin{equation}
\begin{aligned}
M_{1\times N_{E_{rec}}}&=X_{1\times 4}\cdot S_{4\times N_{E_\nu}}\cdot P_{N_{E_\nu}\times N_{E_\nu}}\cdot\sigma_{N_{E_\nu}\times N_{E_\nu}}\cdot R_{N_{E_\nu}\times N_{E_{rec}}}\\
&=X_{1\times 4}\cdot S_{4\times N_{E_\nu}}\cdot P\sigma R_{N_{E_\nu}\times N_{E_{rec}}},
\end{aligned}
\label{matrix_ibd_spec}
\end{equation}
\end{widetext}
where the subscripts denote the dimensions of the corresponding matrices and the number 4 indicates the four isotopes. 
$N_{E_{rec}}$ is the number of bins in the reconstructed prompt energy spectrum while $N_{E_\nu}$ is the number of terms in the discretized sum for integration over $E_\nu$, which is also the number of bins in the extracted isotopic antineutrino spectrum. The ranges of $E_\nu$ and $E_{\rm{rec}}$ are $1.8-10$ MeV and $0.8-10$ MeV, respectively. 
In this study, $N_{E_{rec}}$ was set to 80 based on the limit of the virtual detector energy resolution whereas $N_{E_\nu}$ was optimized to 401 after balancing model performance and computational cost.
The element $X_l$ in $X_{1\times 4}$ can be expressed as 
\begin{equation}
\begin{aligned}
X_l=\sum_{u}^{N_t}\frac{N_p\varepsilon W(t^u)f_l(t^u)}{4\pi L^2 \sum_lf_l(t^u)\epsilon_l}\Delta t\Delta E_\nu\Delta E_{rec},
\end{aligned}
\label{matrix_X}
\end{equation}
where $T_{\text{DAQ}}$ in Eq. ~(\ref{ObsAnt_Eq}) is divided into $N_t$ time units of $\Delta t$, $u$ is the time unit index, and $\Delta E_\nu$ and $\Delta E_{rec}$ are the bin widths of the extracted isotopic antineutrino spectrum and reconstructed prompt energy spectrum, respectively. In Eq.~(\ref{matrix_X}), $W$, $f_l$, and $\epsilon_l$ are reactor-related parameters. $W$ and $f_l$ vary as the reactor evolves while $\epsilon_l$ and the remaining parameters are constants.  $X_l$ is therefore referred to as the reactor dynamic evolution information.

Each row of $S_{4\times N_{E_\nu}}$ represents the binned antineutrino spectrum for isotope $l$, as given by $S_{l}(E_\nu)$. Both $P_{N_{E_\nu}\times N_{E_\nu}}$ and $\sigma_{N_{E_\nu}\times N_{E_\nu}}$ are diagonal matrices whose diagonal elements are given by $P_{ee}(L, E_\nu)$ and $\sigma_{IBD}(E_\nu)$, respectively. The role of $R_{N_{E_\nu}\times N_{E_{rec}}}$ is to map each $E_\nu$ to a spectrum of $E_{rec}$. $R_{N_{E_\nu}\times N_{E_{rec}}}$ is therefore also referred to as the detector response matrix [$R_{qk}$], which is defined as follows:
\begin{equation}
\begin{aligned}
R_{qk}&=R(E^q_\nu,E_{rec}^k)=G(E^q_\nu, \frac{E_{rec}^k+E_{rec}^{k+1}}{2})\\&=G(E^q_\nu, E_{rec}^k+\frac{\Delta E_{rec}}{2}),
\end{aligned}
\label{matrix_R}
\end{equation}
where $q$ is the index for binning $E_\nu$, $q\in[1,~2,~\cdots,~N_{E_\nu}]$, and $k\in[1,~2,~\cdots,~N_{E_{rec}}]$. In contexts that do not involve the oscillation parameters or unfolding, $P_{N_{E_\nu}\times N_{E_\nu}}$, $\sigma_{N_{E_\nu}\times N_{E_\nu}}$, and $R_{N_{E_\nu}\times N_{E_{rec}}}$ can be pre-multiplied to obtain the matrix $P\sigma R_{N_{E_\nu}\times N_{E_{rec}}}$.

The matrix multiplication relation in Eq.~(\ref{matrix_ibd_spec}) provides the mathematical foundation for constructing the FNN architecture presented in Section ~\ref{sec:ffnn}. Furthermore, $X_{1\times 4}$ and $M_{1\times N_{E_{rec}}}$ respectively constitute a sample and its associated target in our dataset, which serve as a feature-label pair for supervised learning in the FNN model implemented in this study.

\begin{figure}[ht]
    \centering
    \includegraphics[width=0.5\textwidth]{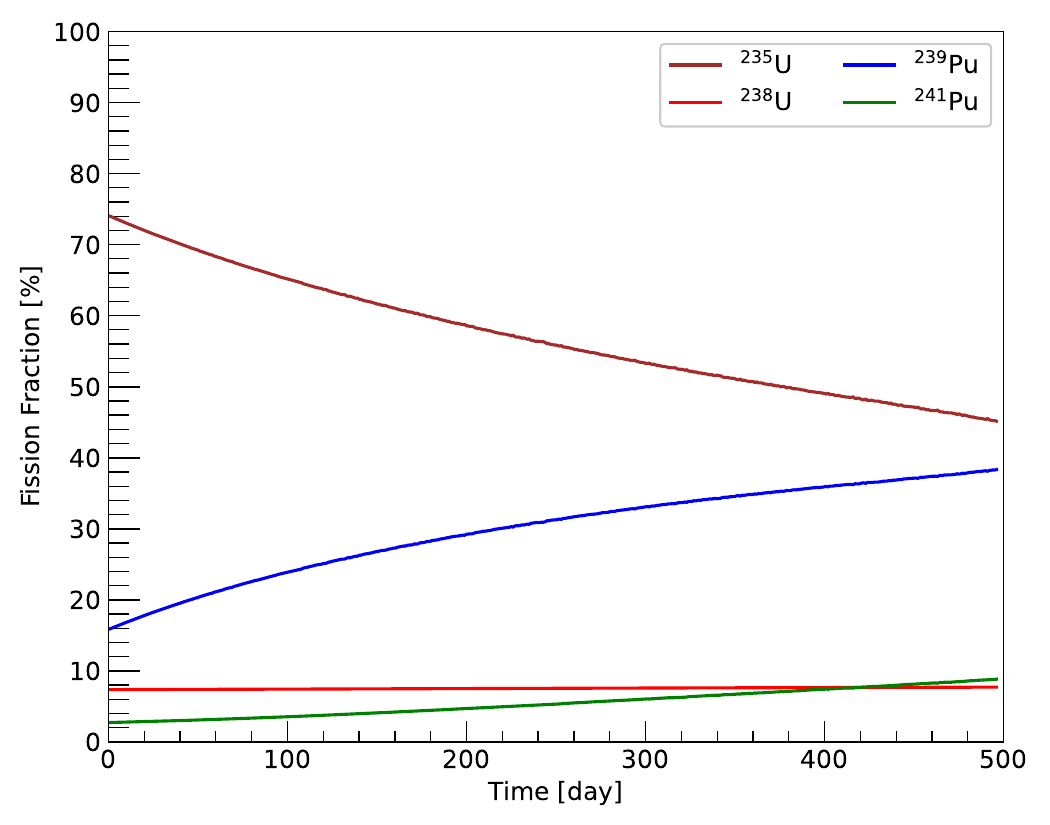}
    \caption{The evolution of fissile fractions for the four main isotopes in the reactor core as a function of operation day, which includes one complete refueling cycle~\cite{DayaBay:2016ssb}. The cumulative fission fraction of the four main isotopes used in this experiment is normalized to unity, and other isotopes contributing less than 0.3\% are excluded from our analysis.    }
    \label{fig:fissionFractionwithtime}
\end{figure}
As described in Eq.~(\ref{matrix_X}), the fission fraction varies dynamically with burn-up as the reactor operates. In each reactor core refueling cycle, the cycle burn-up can be calculated as \cite{DayaBay:2016ssb}
\begin{equation} 
    \text{Burn-up}=\frac{W \cdot D}{M_{\mathrm{U}^{ini}}},
    \label{Eq:Burn-up}
\end{equation}
where $W$, $D$, and ${M_{\mathrm{U}^{ini}}}$ represent the total thermal power of the reactor, the number of days since the refueling cycle started, and the mass of the initial uranium fuel loaded into the reactor, respectively. The unit for burn-up is $\rm{GW_{th}}\cdot day\cdot tonne^{-1}$. Given that the real-time power output of the reactor is dynamic and cannot exceed its maximum capacity of 2.9 GW$_{\rm{th}}$ for safe operation, we used a random number generator for a normal distribution with a mean of 2.9\,GW$_{\rm{th}}$ and downward fluctuation of 0.5\% to determine the daily average power output of the virtual reactor~\cite{Zeng:2023att}. By incorporating the fission fraction evolution data of the isotopes during a complete burn-up cycle from Ref. ~\cite{DayaBay:2016ssb}, we obtained the evolution of the fission fractions for the four main isotopes as a function of the operation day, as shown in figure~\ref{fig:fissionFractionwithtime}.

Under the assumption that the thermal power and fissile fractions for the four main isotopes of the reactor are constant within each day, we accumulated the exposure over each 3-day interval as a sample to create a dataset of 600 simulated samples and their corresponding targets for subsequent analysis.

\section{The implementation of FNN model}
\label{sec:ffnn}
Machine learning algorithms such as neural network (NN) models have attracted increasing attention from high-energy and nuclear physics researchers~\cite{Li:2022tvg,Qian:2021vnh,Zeng:2023att,Shang:2022ntl,He:2023zin,ming2022nuclear}. However, most of these applications are characterized by black-box models in which the meaning of the model parameters are challenging to understand or interpret. In this section, we present a FNN-based white-box model where each layer and parameter has a clear physical or mathematical meaning, thereby ensuring the interpretability of the model.

\subsection{Mathematical foundations of the FNN model}
\label{subsec:model_design}

The NN is a powerful machine learning model that has been widely explored and applied across various fields. The universal approximation theorem~\cite{Cybenko:1989iql,Hornik:1991sec} implies that any continuous function can be approximated with arbitrary precision using an appropriate NN, even if the NN is an FNN with only one hidden layer containing a sufficient number of neurons. However, the internal structure and parameters of the NN in such scenarios often lack physical meaning or interpretability. This results in black-box models, which are not fully trusted by high-energy physicists. Therefore, we designed and implemented a white-box NN model in this study for converting the mathematical mapping function in Eq. (\ref{matrix_ibd_spec}) to a FNN model.

\begin{widetext}
\begin{figure}[ht]
    \centering
    \includegraphics[width=0.98\textwidth]{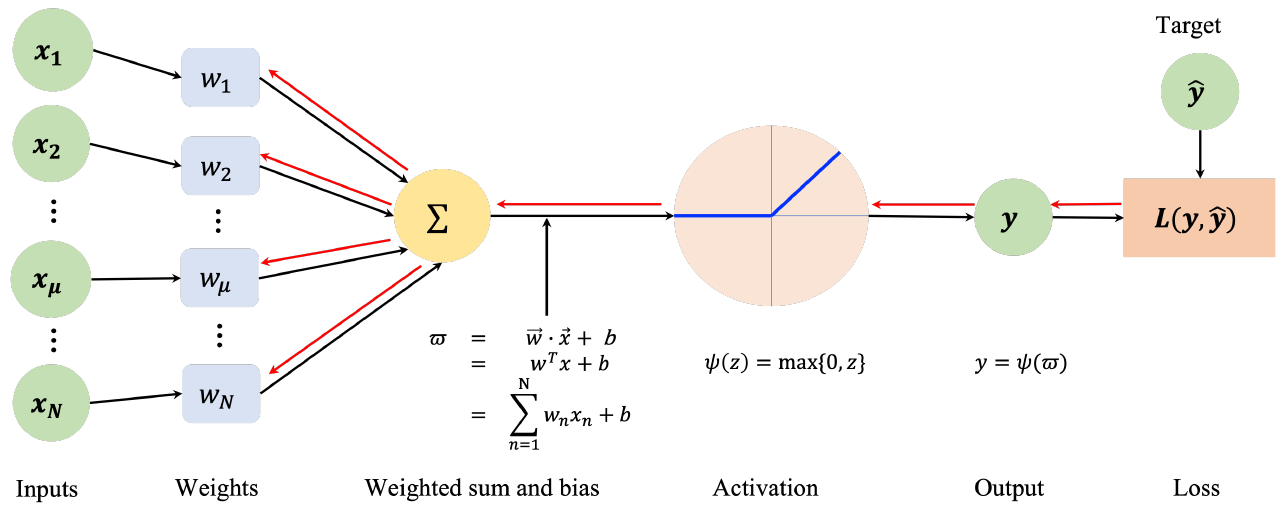}
    \begin{minipage}{\textwidth}
        \caption{An example illustration of the structure of a single-layer perceptron along with forward (black flow arrows) and backward (red flow arrows) propagation.}
        \label{fig:single_layer_perceptron}
    \end{minipage}
   
\end{figure}
\end{widetext}

An FNN is typically composed of one to several single-layer perceptrons, which are considered the fundamental building units of the FNN and play a vital role in its overall functionality~\cite{goodfellow2016deep}. 
Each perceptron in the FNN follows the computational flow shown in figure~\ref{fig:single_layer_perceptron} to process data. Forward and backward propagation are two phases in the NN training process that interact to optimize network performance.

During the forward propagation phase, the perceptron performs computation by computing the dot product of the input vector $\overrightarrow{x}=[x_1,x_2,...,x_N]^T$ with the weight coefficient vector $\overrightarrow{w}=[w_1,w_2,...,w_N]^T$, adding the bias $b$, and applying the activation function $\psi$ to yield the activation result $y$ as the output. The discrepancy between the output $y$ and target $\hat{y}$ is then calculated using the loss function $L(y,\hat{y})$. Forward propagation provides the foundation for evaluating network performance. Backward propagation in turn determines how the network parameters (weights and bias) are updated to reduce loss. It can be described as
\begin{equation}
\begin{aligned}
\omega_\mu'=\omega_\mu-\eta\times[\nabla_{\omega_\mu}L(y,\hat{y})+\lambda\omega_\mu],
\end{aligned}
\label{eq:weight_update}
\end{equation}
\begin{equation}
\begin{aligned}
b'=b-\eta\times\nabla_{b}L(y,\hat{y}),
\end{aligned}
\label{eq:bias_update}
\end{equation}
where $\omega_\mu$ and $\omega_\mu'$ represent the $\mu$-th weight coefficient of the current and subsequent steps, respectively; $b$ and $b'$ the biases of the current and subsequent steps, respectively; and $\eta$ and $\lambda$ are the learning rate and weight decay rate, respectively. This iterative update process of the parameters based on the computed gradients allows the NN to learn and improve its predictions over time.

To allow matrix multiplication in the perceptrons, the bias $b$ must be eliminated, i.e., set to zero. The absence of negative values in our data flow justifies the use of the default rectified linear unit (ReLU) activation function, which is defined as $\psi(z)=max\{0,z\}$. This setup also permits the perceptrons to be chained to perform successive matrix dot product operations, which is integral to the development of our FNN model.

\begin{figure}[ht]
    \centering
    \includegraphics[width=0.5\textwidth]{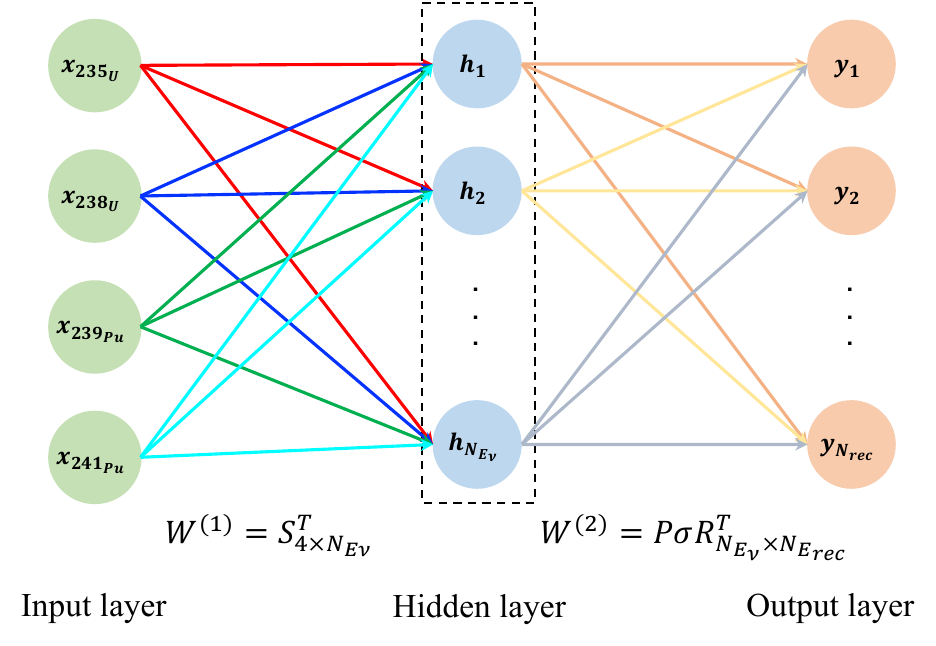}
    \caption{
  The FNN is a white-box model that describes the mapping relation between the reactor dynamic evolution information and reconstructed prompt energy spectrum. The architecture of the FNN model includes an input layer, hidden layer, output layer, and two sets of weight coefficient matrices $W^{(1)}$ and $W^{(2)}$. The weight values between neurons associated with connections of the same color form the rows of the weight coefficient matrix. 
    }
    \label{fig:our_fnn_model}
\end{figure}

As shown in figure~\ref{fig:our_fnn_model}, the architecture of the FNN model consists of three layers comprising, from left to right, the input, hidden, and output layers with four, $N_{E_{\nu}}$, and $N_{E_{rec}}$ neurons, respectively. The neurons in adjacent layers are connected using a fully connected approach; that is, each neuron in one layer is connected to every neuron in the subsequent layer with no connections between neurons within the same layer. The training process of the model starts from the input layer, at which each neuron receives the reactor dynamic evolution information corresponding to its fission isotope. The output of the hidden layer is the scaled total spectrum [$h_q$] of antineutrinos emitted by the reactor. The output layer then provides a predicted reconstructed prompt energy spectrum [$y_k$]. The two weight coefficient matrices $W^{(1)}$ and $W^{(2)}$ correspond to the transposes of the matrices $S_{4\times N_{E_\nu}}$ and $P\sigma R_{N_{E_\nu}\times N_{E_{rec}}}$, respectively.  The matrix $W^{(1)}$ contains the fission isotope antineutrino spectra to be extracted, which are learned during training. In contrast, the matrix $W^{(2)}$ is fixed as $P\sigma R^T_{N_{E_\nu}\times N_{E_{rec}}}$ because it is assumed to be a constant matrix without uncertainties in this study. The FNN is therefore a supervised learning model that iteratively refines $W^{(1)}$ to minimize the discrepancies between its outputs and corresponding targets.

\subsection{Training strategy}
\label{subsec:training_strategy}
 
All the samples generated in Section ~\ref{sebsec:samples} were utilized solely to train the FNN model. The validation and testing processes were omitted. This approach was chosen because our aim is to minimize the loss function during the training process to determine the optimal $W^{(1)}$ for extracting the four main isotopic antineutrino spectra. Our focus is on optimizing spectra extraction performance rather than evaluating model performance across various datasets, as well as on simplifying the process and aligning with our primary research objective.

The loss function is a fundamental component in deep-learning models. 
It serves as the criterion for evaluating how well the model predictions match the actual outcomes and provides a numerical indicator of model accuracy. 
The Combined Neyman–Pearson (CNP) chi-square model is a statistical model frequently employed in HEP experiments to quantify the error between predicted and measured values~\cite{Ji:2019yca}. Based on this model, we define the loss function for the FNN model as 
\begin{equation}
\begin{aligned}
\chi^2_{CNP}=\sum_{k=1}^{N_{E_{rec}}}\frac{[M_k-y_k(W^{(1)})]^2}{3/\left[\frac{1}{M_k}+\frac{2}{y_k(W^{(1)})}\right]},
\end{aligned}
\label{eq:CNP_Eq}
\end{equation}
where $M_k$ is the IBD event number in the $k$-th bin for the measured reconstructed prompt neutrino energy spectrum given by Eq.~(\ref{ObsAnt_Eq}) and $y_k$ is the corresponding predicted value output of the model. We used this loss function to guide the optimization process of $W^{(1)}$ during the training process so that the FNN was driven towards increasingly precise predictions. 

After defining the loss function, it is essential to select a suitable optimizer, learning rate schedule, batch size, and epoch, among other hyperparameters. Following hyperparameter tuning using the Optuna framework~\cite{akiba2019optuna} and extensive testing,  we developed two training strategies denoted as the short- and long-epoch strategies to investigate the performance of the FNN model in extracting the antineutrino spectra of the four fission isotopes from the reconstructed prompt energy spectrum. As shown in Table ~\ref{tab:schemes}, a critical commonality between these two strategies is the segmentation of the hidden layer in the FNN model into multiple partitions or parallel hidden layers. This setup allows distinct learning and weight decay rates to be assigned to each partition to facilitate differential performance outcomes. Because the focus in this study is not on the isotope antineutrino spectra above 8 MeV, i.e., in the (303, 401] partition or the matrix $P\sigma R_{N_{E_\nu}\times N_{E_{rec}}}$, we fixed their learning and weight decay rates to zero and disabled the gradient calculations for the corresponding weight coefficients. Additionally, we set the initial value of $W^{(1)}$ based on the Huber-Mueller model.
\begin{table}[!htb]
\centering
\caption{Configurations of the two training strategies for FNN model.
The configurations were derived based on our empirical knowledge and optimized using Optuna~\cite{akiba2019optuna}. The partition numbers correspond to neuron indices in the hidden layer of the FNN model. ReduceLROnPlateau is a Python class that dynamically adjusts the learning rate during deep learning model training to improve convergence speed and performance~\cite{ReduceLROnPlateau}.}

\label{tab:schemes}
\begin{tabular*}{8.5cm} {@{\extracolsep{\fill} } c|c|c}
\toprule
Strategy & Short-epoch & Long-epoch \\ \hline
Epoch & 2e3 & 2e6 \\ \hline
Optimizer & AdamW & Adam \\ \hline
\tabincell{c}{The partitions of\\ the hidden layer}&\multicolumn{2}{c}{\tabincell{c}{[1], (1, 180], (180, 225], (225, 303], (303, 401]}} \\ \hline
\tabincell{c}{The learning\\ rates for the\\ hidden layer}&\multicolumn{2}{c}{\tabincell{c}{[3.4892e-4, 9.9485e-4, 2.754e-4, 1.8272e-4, 0]}} \\ \hline
\tabincell{c}{The weight\\ decay rates for\\ the hidden layer}&\tabincell{c}{[7.418e-3, 7.748e-3,\\ 4.155e-3, 9.999e-3, 0]}&[0, 0, 0, 0, 0] \\ \hline
\tabincell{c}{The learning\\ rate for the\\ output layer} & \multicolumn{2}{c}{0} \\ \hline
\tabincell{c}{The weight\\ decay for the\\ output layer} & \multicolumn{2}{c}{0} \\ \hline
\tabincell{c}{Learning rate\\ scheduler} & \tabincell{c}{ReduceLROnPlateau \\($\rm{factor=0.32}$,\\ $\rm{patience=1e2}$)} & \tabincell{c}{ReduceLROnPlateau\\ ($\rm{factor=0.32}$,\\ $\rm{patience=1e4}$)\\ \& $\rm{epoch\geq 2e5}$ }\\ \hline
Batch size & \multicolumn{2}{c}{30} \\ 
\bottomrule
\end{tabular*}
\end{table}

As indicated by their names, the main distinction between the short- and long-epoch strategies lies in the epochs. The short-epoch strategy leverages the AdamW~\cite{Loshchilov2017DecoupledWD} optimizer with non-zero weight decay rates for faster loss reduction. In contrast, in the long-epoch strategy, the Adam~\cite{Kingma2014AdamAM} optimizer is applied without weight decay, i.e., the weight decay rates are set to zero. Superior convergence results were obtained using the long-epoch strategy. The results are presented and discussed in Section ~\ref{sec:analysis&results}. As illustrated in Table ~\ref{tab:schemes}, these circumstances also led to minor differences in the configurations of the learning rate schedulers. Nonetheless, the same metric, i.e., the sum of the losses for all samples denoted as $\chi^2_{\sum CNP}$, was monitored in both schedulers.

We also extracted the antineutrino spectra of the four fission isotopes using the $\chi^2$ minimization method to provide a comparison and benchmark for the FNN model. We employed the Minuit2 minimization library from ROOT~\cite{Brun:1997pa} to implement this method. $\chi^2_{\sum CNP}$ was used as the objective function to be minimized to find the best fit. The same dataset as that for the FNN model was used as the measured value in this fitting process. In contrast, the predicted value was derived from Eq.~(\ref{matrix_ibd_spec}) where the $S_{4\times N_{E_\nu}}$ matrix elements corresponding to $\leq 8$ MeV are the parameters to be fitted and the remaining elements considered as fixed parameters in the fitting procedure. We adopted the ``Combined" minimizer algorithm to minimize the objective function with initial fitting values from the Huber-Mueller model and fitting step sizes of 1\% of the order of magnitude of these values. We set the tolerance for the fitting procedure to $1\times 10^{-30}$. The fitting stopped automatically only when the improvement in the $\chi^2_{\sum CNP}$ value between consecutive iterations fell below this threshold.


The FNN model was implemented using PyTorch~\cite{paszke2019pytorch}, a Python-based deep learning library that supports both CPU and GPU platforms and is one of the mainstream tools for developing and training NN models. A NVIDIA GeForce RTX 3060 Ti GPU platform was used to deploy the FNN model, whereas tasks involving Optuna and ROOT were performed on two identical servers, each of which was equipped with two 28-core Intel(R) Xeon(R) Gold 6330 CPUs @ 2.00 GHz.

\section{Results and discussions}
\label{sec:analysis&results}
\begin{figure}[ht]
    \centering
    \includegraphics[width=0.5\textwidth]{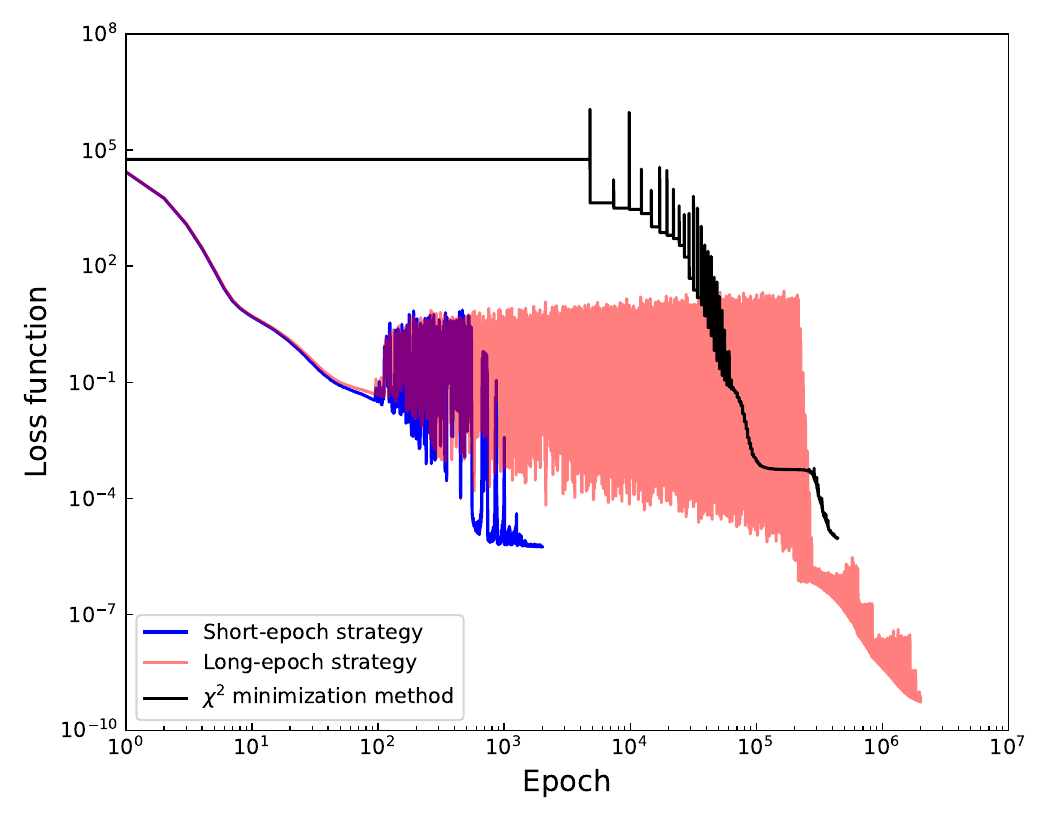}
    \caption{Evolution of loss function across epochs for the short- and long-epoch strategies and the $\chi^2$ minimization method. The epochs of the first two were manually specified to be 2e3 and 2e6, respectively, while that of the $\chi^2$ minimization method was automatically determined as approximately 4.39e5. 
    }
    \label{fig:three_losses}
\end{figure}

To facilitate the discussion and comparative analysis of the short- and long-epoch strategies of our FNN model and the $\chi^2$ minimization method, we first consider their performance in fitting all the samples and reducing the losses. As shown in figure~\ref{fig:three_losses}, did the loss $\chi^2_{\sum CNP}$ decreased more rapidly in both FNN strategies, and lower ultimate $\chi^2_{\sum CNP}$ values were obtained compared to those obtained by the $\chi^2$ minimization method. The $\chi^2_{\sum CNP}$ values at the conclusion of the epochs are $5.51\times10^{-6}$, $5.42\times10^{-10}$, and $9.34\times10^{-6}$, respectively. 

The short-epoch strategy can rapidly reduce the loss in the early stages of training mainly because of the regularization effects and optimization efficiency due to the combination of nonzero weight decay rates and the AdamW optimizer. However, in the later stages of training, the model must be able to respond to small changes in the loss function for fine adjustments of the parameters. Weight decay may interfere with this process and make it challenging for the model to determine the optimal solution within regions of small loss function gradients. 

\begin{figure}[ht]
    \centering
    \includegraphics[width=0.5\textwidth]{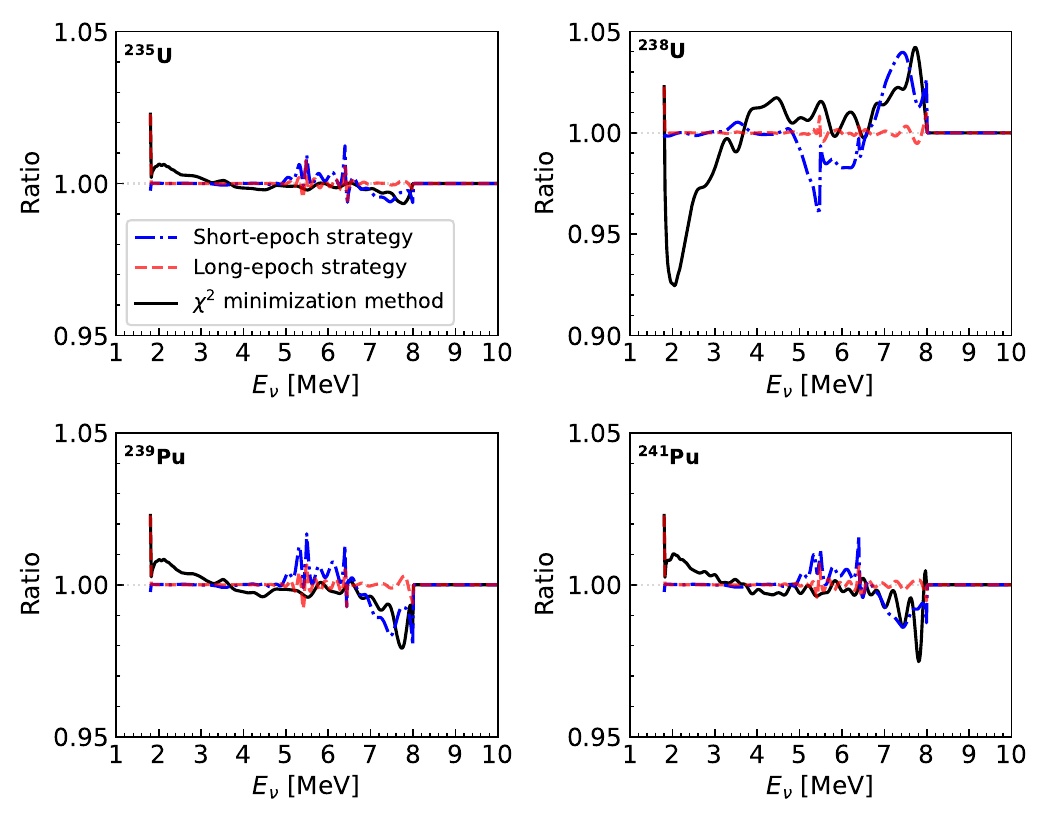}
        \caption{Comparison of the ratios between the four isotope antineutrino spectra extracted using the short- and long-epoch strategies in our FNN model and the $\chi^2$ minimization method, and the assumed true spectra described by Eq.~(\ref{eq:raa_iso_antineu_spectrum}).
    } 
    \label{fig:isotopic_three_ratios}
\end{figure}

Figure~\ref{fig:isotopic_three_ratios} shows a comparison of the performance in extracting the antineutrino spectra of the four isotopes using these three approaches. The extraction performance decreases in the order of the long-epoch strategy, short-epoch strategy, and $\chi^2$ minimization method. The FNN model accurately extracted the antineutrino spectra of $\mathrm{^{235}U}$, $\mathrm{^{239}Pu}$, and $\mathrm{^{241}Pu}$ in the energy range of $2-5$ MeV. The FNN model with the short-epoch strategy achieved relative errors of less than 2\% in the $5-8$ MeV range, which decreased to less than 1\% with the long-epoch strategy. In comparison, the $\chi^2$ minimization method achieved relative extraction errors of less than 2\% and 3\% for these three isotopes in the respective energy ranges. For the isotope $\mathrm{^{238}U}$, both the short-epoch strategy and $\chi^2$ minimization method showed relatively poor extraction performance compared to that for the other isotopes. The maximum extraction relative errors in the $2-8$ MeV range are approximately 4\% and 8\%, respectively, whereas only the long-epoch strategy maintained relative errors of less than 1\%.

It is worth noting that although $\mathrm{^{241}Pu}$ has a lower average fission fraction throughout the entire refueling cycle compared to $\mathrm{^{238}U}$, the extraction performance for the former is better in all the extraction approaches. This indicates that in addition to large fission fractions, significant variations are also crucial for extracting isotopic antineutrino spectra accurately. Greater variations produce better extraction results. This is further confirmed by the extraction performance for the $\mathrm{^{235}U}$ and $\mathrm{^{239}Pu}$ antineutrino spectra. Therefore, such long epochs are employed in the long-epoch strategy primarily to enhance the extraction performance for $\mathrm{^{238}U}$. Overall, regardless of the extraction approach used, the extraction performance for the isotopic antineutrino spectra in descending order is as follows: $\mathrm{^{235}U}$, $\mathrm{^{239}Pu}$, $\mathrm{^{241}Pu}$, and $\mathrm{^{238}U}$.

The above results and discussion reveal that because of the exceptional capability of NNs in optimizing large-scale parameters, the FNN model achieved faster and more effective convergence than the traditional $\chi^2$ minimization method. Based on PyTorch's extensive array of optimization algorithms~\cite{Kim_pytorch_optimizer_optimizer_2021}, various model training strategies can be designed to satisfy the practical requirements for extracting isotope antineutrino spectra. Moreover, executing spectrum extraction algorithms on GPU platforms can significantly increase the inference speed of the process, thereby improving extraction efficiency.

\section{Summary and outlook}
\label{sec:summary&outlook}
In this study, we presented an FNN model designed to infer and extract the corresponding antineutrino spectra generated by the fission of $\mathrm{^{235}U}$, $\mathrm{^{238}U}$, $\mathrm{^{239}Pu}$, and $\mathrm{^{241}Pu}$ from the reconstructed prompt energy spectrum measured by the detector in a reactor neutrino experiment. Using a simulated short-baseline reactor neutrino experiment with an exposure of $(2.9\times 5\times 1800)~\rm{GW_{th}\cdot tonnes\cdot days}$, we demonstrated how this FNN model establishes a mapping from reactor evolution information to the reconstructed prompt energy spectrum and enables the extraction of antineutrino spectra for the four isotopes through its training process. 

By comparing the extraction effects of the short- and long-epoch training strategies for our FNN model with the traditional $\chi^2$ minimization method, as shown in figure~\ref{fig:isotopic_three_ratios}, we found that the FNN model converged faster and better, and the performance of the three approaches for extracting the isotope antineutrino spectra in descending order is as follows: long-epoch strategy, short-epoch strategy, and $\chi^2$ minimization method. Furthermore, the relative extraction errors of the antineutrino spectra for the four isotopes are reduced to less than 1\% in the $2-8$ MeV range of interest by the FNN model with the long-epoch strategy, which is better than the error of 8\% or less obtained using the $\chi^2$ minimization method in the control group. These results show that the FNN model has considerable potential for extracting fission isotope antineutrino spectra.

In the near future, TAO will serve as a satellite experiment of JUNO and achieve an energy resolution exceeding 2\% at 1 MeV in measuring reactor antineutrinos~\cite{JUNO:2020ijm}. Its primary physics goals include constraining the fine structures of isotope antineutrino spectra and providing a model-independent reference spectrum for JUNO and a benchmark measurement to test nuclear databases. Employing the FNN model in high-precision experiments such as TAO would therefore be an excellent match. In addition, depending on the research objectives, new NN models can be developed using the methodologies outlined in this study to further investigate a broader range of physics topics such as unfolding, neutrino oscillation parameter measurements, sterile neutrino searches, and reactor monitoring. For example, the unfolded neutrino energy spectrum is represented by the output of the hidden layers in our FNN model, which can achieve a relative error of less than 1\% in the $2-8$ MeV range.

\end{document}